\documentclass[hyper]{JHEP} 

\usepackage{epsfig}

\newcommand\fverb{\setbox\pippobox=\hbox\bgroup\verb}

\newcommand\fverbdo{\egroup\medskip\noindent%

            \fbox{\unhbox\pippobox}\ }

\newcommand\fverbit{\egroup\item[\fbox{\unhbox\pippobox}]}

\newbox\pippobox

\title{Note About Equivalence of
$F(\tilde{R})$ and Scalar Tensor
Ho\v{r}ava-Lifshitz Gravities}
\author{J. Kluso\v{n}\\
Department of
Theoretical Physics and Astrophysics\\
Faculty of Science, Masaryk University\\
Kotl\'{a}\v{r}sk\'{a} 2, 611 37, Brno\\
Czech Republic\\
E-mail: \email{klu@physics.muni.cz}}
\preprint{}

 \abstract{In this  note we study the relation
 between  $F(\tilde{R})$  and    scalar tensor
   Ho\v{r}ava-Lifshitz
 gravity. We find that
due to the  broken
diffeomorphism invariance
 corresponding scalar tensor theory
has more complicated form than in case
of the full diffeomorphism invariant
$F(R)$ theory of gravity. We also show
that in the low energy limit this
theory flows to the relativistic scalar
tensor theory of gravity.}


\def\tR{\tilde{R}}

\def\be{\begin{equation}}

\def\ee{\end{equation}}

\def\bea{\begin{eqnarray}}

\def\eea{\end{eqnarray}}

\def\bx{\mathbf{x}}

\newcommand{\hg}{\hat{g}}

\newcommand{\mG}{\mathcal{G}}

\def\mV{\mathcal{V}}

\newcommand{\mL}{\mathcal{L}}

\begin{document}
\section{Introduction and Summary}\label{first}
In 2009 Petr Ho\v{r}ava formulated new
proposal of quantum theory of gravity
(now known as Ho\v{r}ava-Lifshitz
gravity (HL gravity) that is power
counting renormalizable
\cite{Horava:2009uw,Horava:2008ih,Horava:2008jf}
that is also expected that it reduces
do General Relativity in the infrared
(IR) limit \footnote{For review and
extensive list of references, see
\cite{Horava:2011gd,Padilla:2010ge,Mukohyama:2010xz,Weinfurtner:2010hz}.}.
The HL gravity is based on
 an idea that
the Lorentz symmetry is restored in IR
limit of given theory while it is
absent in its  high energy regime. For
that reason
  Ho\v{r}ava considered
systems whose scaling at short
distances exhibits a strong anisotropy
between space and time,
\begin{equation}
\bx' =l \bx \ , \quad t' =l^{z} t \ .
\end{equation}
In $(D+1)$ dimensional space-time in
order to have power counting
renormalizable theory  requires that
$z\geq D$. It turns out however that
the symmetry group of given theory is
reduced from the full diffeomorphism
invariance of General Relativity  to
the foliation preserving diffeomorphism
\begin{equation}\label{fpdi}
x'^i=x^i+\zeta^ i(t,\bx) \ , \quad
t'=t+f(t) \ .
\end{equation}
The HL gravity was then generalized to
the case of $F(\tR)$ HL gravities in
series of papers
\cite{Chaichian:2010yi,Carloni:2010nx}
\footnote{For further study in given
direction, see
\cite{Chaichian:2010zn,Kluson:2009rk,Kluson:2009xx,Kluson:2010xx},
and for review, see
\cite{Nojiri:2010wj}.}. $F(\tR)$ HL
gravity can be considered as natural
generalization of covariant $F(R)$
gravity.  Current interest to $F(R)$
gravity is caused by several important
reasons. First of all, it is known that
such theory may give the unified
description of the early-time inflation
and late-time acceleration (for a
review, see
\cite{Nojiri:2010wj,Nojiri:2006ri}.)
Moreover, the whole sequence of the
universe evolution epochs: Inflation,
radiation/matter dominance and dark
energy may be obtained within such
theory. The remaining freedom in the
choice of $F(R)$ function could be used
for fitting the theory with
observational data. Second, it is known
that higher derivatives gravity (like
$R^2$-gravity, for a review, see
\cite{Buchbinder:1992rb}) has better
ultraviolet behavior than conventional
General Relativity. Third, modified
gravity is pretending also to be the
gravitational alternative for Dark
Matter. Fourth, it is expected that
consistent quantum gravity emerging
from string/M-theory should be
different from General Relativity.
Hence, it should be modified by
fundamental theory. Of course, all
these reasons remain to be the same
also for the HL gravity.

It is well known  that
the  $F(R)$ gravity is
equivalent to the scalar tensor theory
of gravity \footnote{For review and
extensive list of references, see
\cite{Capozziello:2009nq,Nojiri:2006ri,Buchbinder:1992rb,Nojiri:2010wj}.}.
The scalar tensor theory of gravity
 has relatively simple form
corresponding to the  General Relativity action
coupled with the scalar field with
specified form of the potential term.
In particular, the presence of the
additional scalar mode in $F(R)$ theory of
gravity is clearly seen
in the scalar tensor description of the
$F(R)$ theory of gravity. Further, the
properties of this scalar mode can be
transparently studied in this formulation
as well. On the other hand it turns out
that it is sometimes useful to use the
original form of $F(R)$ theories of
gravity, as for example for the
analysis of the cosmological solutions.

Successes of the equivalence between
$F(R)$ gravity and scalar tensor theory
of gravity naturally implies the
question whether there exists similar
equivalence between $F(\tR)$ HL gravity
and scalar tensor version of the HL
gravity. Examples  of the scalar tensor
HL gravities were introduced in
\cite{Calcagni:2009ar,Kiritsis:2009sh,
Capasso:2009fh,Suyama:2009vy,Romero:2009qs}
where the authors analyzed the
cosmological consequences of HL
gravities \footnote{For further study,
see
\cite{Wang:2010mw,Wang:2009azb,Borzou:2011mz}.}.
The goal of this note is to understand
the relation between these scalar
tensor HL gravities and $F(\tR)$
gravities.
 Our
procedure is similar as in case of
$F(R)$ theories of gravity. We start with
the $F(\tR)$ HL gravity action and
introduce two auxiliary scalar fields
in order to rewrite it into Jordan-like
form. Then we use the anisotropic conformal
transformation of the metric components
in order to map this form of the action
 to the action where
the kinetic term has the canonical
form. By canonical form of the kinetic
term we mean that it has the same form
as the kinetic term in HL gravity that
 is formulated in ADM formalism
\cite{Arnowitt:1962hi}, for review, see
\cite{Gourgoulhon:2007ue,Dengiz:2011aa}.
Now due to the fact that $F(\tR)$
gravity is not fully diffeomorphism
invariant we find that the resulting
theory takes more general form of  the
scalar tensor theory. We also  show
that this theory flows to the
relativistic  scalar tensor theory of
gravity in the low energy limit.

We hope that our result can be useful
for further analysis of the properties
of $F(\tR)$ HL gravities. For example,
the scalar tensor form of $F(\tR)$
theory can be useful for
 the analysis of the fluctuations
around cosmological solutions of $F(\tR)$ HL
 gravities. We hope
to return to these problems in future.

This paper is organized as follows. In
the next section (\ref{second}) we
introduce $F(\tR)$ HL theories of
gravity and map them to generalized
scalar tensor theories of gravity. We
also demonstrate that these theories
flow to standard scalar tensor theories
of gravity in its low energy region. In
Appendix (\ref{Appendix}) we review the
standard equivalence between $F(R)$
theory of gravity and scalar tensor
theory of gravity. We perform this
analysis in the ADM formalism in order
to compare this result with the
analysis performed in the main body of
the paper.
\section{$F(\tR)$  HL Gravity in
Einstein Frame}\label{second}
We begin this section with the review
of basic properties of $F(\tR)$ HL gravity.
Our convention is as follows.
We consider $D+1$ dimensional
manifold $\mathcal{M}$ with the
coordinates $x^\mu \ , \mu=0,\dots,D$
and where $x^\mu=(t,\bx) \ ,
\bx=(x^1,\dots,x^D)$. We presume that
this space-time is endowed with the
metric $\hat{g}_{\mu\nu}(x^\rho)$ with
signature $(-,+,\dots,+)$. Suppose that
$ \mathcal{M}$ can be foliated by a
family of space-like surfaces
$\Sigma_t$ defined by $t=x^0$. Let
$g_{ij}, i,j=1,\dots,D$ denotes the
metric on $\Sigma_t$ with inverse
$g^{ij}$ so that $g_{ij}g^{jk}=
\delta_i^k$. We further introduce the operator
$\nabla_i$ that is covariant derivative
defined with the metric $g_{ij}$.
We introduce the future-pointing unit normal vector $n^\mu$ to the surface $\Sigma_t$.
In ADM variables one has $n^0=\sqrt{-\hat{g}^{00}}$,
$n^i=-\hat{g}^{0i}/\sqrt{-\hat{g}^{00}}$.
 We also define  the lapse
function $N=1/\sqrt{-\hat{g}^{00}}$ and
the shift function
$N^i=-\hat{g}^{0i}/\hat{g}^{00}$. In
terms of these variables we write the
components of the metric
$\hat{g}_{\mu\nu}$ as
\begin{eqnarray}
\hat{g}_{00}=-N^2+N_i g^{ij}N_j \ ,
\quad \hat{g}_{0i}=N_i \ , \quad
\hat{g}_{ij}=g_{ij} \ ,
\nonumber \\
\hat{g}^{00}=-\frac{1}{N^2} \ , \quad
\hat{g}^{0i}=\frac{N^i}{N^2} \ , \quad
\hat{g}^{ij}=g^{ij}-\frac{N^i N^j}{N^2}
\ .
\nonumber \\
\end{eqnarray}
We further define the extrinsic
derivative
\begin{equation}
K_{ij}=\frac{1}{2N}
(\partial_t g_{ij}-\nabla_i N_j-
\nabla_j N_i) \ .
\end{equation}
The general formulation of
Ho\v{r}ava-Lifshitz $F(\tR)$ gravity
was presented in series of papers in
\cite{Chaichian:2010yi,Carloni:2010nx}
\footnote{For further study in given
direction, see
\cite{Chaichian:2010zn,Kluson:2010xx},
and for review, see
\cite{Nojiri:2010wj}}. The action
introduced in \cite{Chaichian:2010yi}
takes the form
\begin{equation}
\label{actionNOJI} S_{F(\tilde{R})}=
\zeta^2\int dt d^D\bx \sqrt{g}N F
(\tilde{R}) \, ,
\end{equation}
where
\begin{equation}
\label{tR} \tilde{R}= K_{ij}\mG^{ijkl}K_{kl}
+ \frac{2\mu}{\sqrt{-\hat{g}}} \partial_\mu \left(\sqrt{-\hat{g}}n^\mu K\right)
 -\frac{2\mu}{\sqrt{g}N} \partial_i \left(\sqrt{g}g^{ij}\partial_j N\right)
 -\mathcal{V}(g) \, ,
\end{equation}
where $\mu$ is constant, $K=K_{ij}g^{ji}$ and where  the
generalized de Witt metric $\mG^{ijkl}$  is defined as
\begin{equation}
\mG^{ijkl}=\frac{1}{2}(g^{ik}g^{jl} + g^{il}g^{jk})-\lambda g^{ij}g^{kl} \, ,
\end{equation}
where $\lambda$ is real constant that
is believed that it flows to  $1$ in
its low energy limit. More precisely we
 presume that $F(R)$
gravity is recovered in the limit
$\lambda\rightarrow 1, \mu\rightarrow
1$ and $\zeta^2\rightarrow (16\pi
G)^2$.
 Finally $\mV(g)$ depends on   $g_{ij}$ and its
covariant derivatives
 whose explicit
form  was suggested in
\cite{Sotiriou:2009bx}.
%
Our goal is to map the action
(\ref{actionNOJI}) to the scalar tensor
form of HL gravity. The first step is
to rewrite the action
(\ref{actionNOJI}) into an equivalent
form
\begin{eqnarray}\label{SFtR}
S_{F(\tilde{R})}
=\zeta^2\int dt d^D\bx \left( \sqrt{g}N
B( K_{ij}\mG^{ijkl}K_{kl}
-\mV(g)-A)+\nonumber \right.
\\
\left. +\sqrt{g}N F(A) -2\mu
\sqrt{g}N\nabla_n B
K  + 2\mu
\partial_i B \sqrt{g}g^{ij}
\partial_j N \right) \ , \nonumber \\
\end{eqnarray}
where
 \begin{equation} \nabla_n X=
 \frac{1}{N}(\partial_t X-N^i\partial_i
 X) \ .
 \end{equation}
 Then from (\ref{SFtR}) we find the
equation of motion for $A$
\begin{equation}
-B+F'(A)=0 \ .
\end{equation}
 Assuming an existence of
  the inverse function $\Psi$
 defined as $\Psi (F')(A)=A$ we would be able to
 determine $A$
 as a function of $B$
 \begin{equation}
 A=\Psi (B) \ .
 \end{equation}
 Inserting this result into the action (\ref{SFtR})
we obtain
\begin{eqnarray}\label{SFtR2}
S_{F(\tilde{R})}
&=&\zeta^2\int dt d^D\bx \left(
\sqrt{g}N B( K_{ij}\mG^{ijkl}K_{kl}
-\mV(g))-\sqrt{g}NV(B)-\nonumber
\right.
\\
& &\left.  -2\mu
\sqrt{g}N\nabla_n B
K  + 2\mu
\partial_i B \sqrt{g}g^{ij}
\partial_j N \right) \ , \nonumber \\
\end{eqnarray}
where
\begin{equation}\label{VB}
V(B)=B\Psi(B)-F(\Psi(B)) \ .
\end{equation}
 Let us now consider following
 anisotropic Weyl transformation
\begin{equation}\label{anWeyl}
N'=\Omega^\omega N \ , \quad N'_i=\Omega^2
N_i \ , \quad g_{ij}=\Omega^2 g_{ij} \
,
\end{equation}
where $\omega$ is free parameter whose
value will be specified below. It is easy
to see  that the spatial
connection
\begin{equation}
\Gamma_{ij}^k= \frac{1}{2}g^{kl}
(\partial_ i g_{lj}+\partial_j g_{li}
-\partial_l g_{ij})
\end{equation}
 transforms under (\ref{anWeyl})
 as
\begin{eqnarray}
\Gamma'^k_{ij}= \Gamma^k_{ij}+
\frac{1}{\Omega} (\delta^k_i\partial_j
\Omega+\delta^k_j\partial_i
\Omega-g^{kl}
\partial_l \Omega g_{ij})
\nonumber \\
\end{eqnarray}
and the extrinsic curvature transforms
as
\begin{eqnarray}
K'_{ij}=
\Omega^{2-\omega} K_{ij}+
\Omega^{1-\omega}\nabla_n \Omega g_{ij} \ .
\nonumber \\
\end{eqnarray}
In the same way we find
\begin{eqnarray}
\nabla_n' B=\frac{1}{\Omega^\omega} \nabla_n
B \ , \quad
K'=\frac{K}{\Omega^\omega}+D\frac{\nabla_n
\Omega}{\Omega^{1+\omega}} \ . \nonumber \\
\end{eqnarray}
Then it is easy to see that the
kinetic part of the action
(\ref{SFtR2}) transforms as
\begin{eqnarray}\label{SFtR2kin}
&&\zeta^2 \int dt d^D\bx \sqrt{g }N [ B
K_{ij}\mG^{ijkl}K_{kl}- 2\mu
 K \nabla_n B +
\frac{2\mu}{N} g^{ij}\partial_j B
\partial_i N]
\rightarrow \nonumber \\
& &\zeta^2 \int dt
d^D\bx\sqrt{g}N\Omega^{D+\omega}B\left[
\frac{1}{\Omega^{2\omega}}
 K_{ij}\mG^{ijkl}K_{kl}
+\frac{2}{\Omega^{1+2\omega}} (1-\lambda
D)K\nabla_n\Omega +\right.\nonumber
\\
&+& \left. \frac{1}{\Omega^{2+2\omega}}
\nabla_n\Omega\nabla_n\Omega (1-\lambda
D)D-2\mu \frac{1}{\Omega^{2\omega}B}\nabla_n
BK-2\mu\frac{1}{\Omega^{1+2\omega}B}
D\nabla_n B\nabla_n\Omega+\right.\nonumber \\
&+& \left. \frac{2\omega\mu}{B\Omega^3 N}\partial_i B
g^{ij}\partial_j\Omega+
\frac{2\mu}{B\Omega^2N}
\partial_i B g^{ij}\partial_jN
 \right] \ .
\nonumber \\
\end{eqnarray}
Our goal is to choose $\Omega$ in such
a way so that the kinetic term takes
the canonical form as in the  scalar
tensor HL gravity. The requirement
implies following relation between
$\Omega$ and $B$
\begin{equation}\label{OmegaB}
\Omega=B^{\frac{1}{\omega-D}} \
\end{equation}
With such a form of $\Omega$ the
expression (\ref{SFtR2kin})
simplifies as
\begin{eqnarray}
& &\zeta^2 \int dt
d^D\bx\sqrt{g}N\left[
 K_{ij}\mG^{ijkl}K_{kl}
+\frac{2}{B(\omega-D)} (1-\lambda
D)K\nabla_n B +\right.\nonumber
\\
&+&\frac{(1-\lambda
D)D}{(\omega-D)^2B^2} \nabla_n
B\nabla_n B -\frac{2\mu}{B} \nabla_n
BK-2\mu\frac{D}{(\omega-D)B^2}
\nabla_n B\nabla_n B+\nonumber \\
&+& \left.\frac{2\omega\mu}{(\omega-D)
N}B^{\frac{2D-2}{\omega-D}}\partial_i B
g^{ij}\partial_jB+ \frac{2\mu}{N}
B^{\frac{D+\omega-2}{\omega-D}}\partial_i
B g^{ij}\partial_jN \right] \ .
\nonumber \\
\end{eqnarray}
Note that this expression is not well
defined for $\omega=z=D$ where $z$ is the
scaling dimension \cite{Horava:2009uw}. This follows
from the fact that the kinetic term of the
HL gravity is invariant under anisotropic
scaling transformation when $\omega=D$.
Now we come to the analysis of the
potential term. We consider the SVW
potential term \footnote{For simplicity
we consider the potential term without
cosmological constant contribution.}
\begin{eqnarray}\label{mV}
\mV(g)&=&
 g_1 R+
\frac{1}{\zeta^2}(g_2 R^2+ g_3
R_{ij}R^{ij})+\nonumber \\
&+&\frac{1}{\zeta^4} (g_4 R^3+g_5 R
R_{ij}R^{ij}+g_6 R^i_j R^j_k R^k_i)+
\nonumber \\
&+&\frac{1}{\zeta^4} [g_7 R \nabla^2 R
+g_8(\nabla_i R_{jk}) (\nabla^i
R^{jk})]+\dots \ , \nonumber \\
\end{eqnarray}
where the coupling constants $g_s,
(s=0,1,2,\dots)$ are dimensionless and
$\dots$ corresponds to the higher order
terms corresponding to the fact that
the critical dimension of
$D-$dimensional HL gravity is $z=D$.
 The relativistic limit in the IR
requires $g_1=-1$ and $\zeta^2= (16\pi
G)^{-2}$.
 We note that under
transformations (\ref{anWeyl}) the
components of $D-$dimensional
 Ricci
tensor transforms as
\begin{eqnarray}\label{RijOmega}
R'_{ij}&=&  R_{ij}
+2(D-2) \frac{1}{\Omega^2} (\nabla_i
\Omega)(\nabla_j \Omega)-(D-2)
\frac{1}{\Omega}\nabla_i\nabla_j\Omega+
\nonumber \\
&+&(3-D)g_{ij} \frac{\nabla_k\Omega
\nabla^k\Omega}{\Omega^2} -g_{ij}
\frac{\nabla_k\nabla^k\Omega}{\Omega} \
 \nonumber \\
\end{eqnarray}
while $D-$dimensional  scalar curvature transforms as
\begin{equation}\label{ROmega}
 R'= \Omega^{-2} \left(
R-2(D-1)g^{ij}\frac{\nabla_i\nabla_j\Omega}{\Omega}
+(D-1)(4-D)\frac{\nabla_i\Omega\nabla^i\Omega}{\Omega^2}
\right) \ .
\end{equation}
To proceed further it is useful  to
separate the contribution proportional
to $R$
in (\ref{mV}) so that we rewrite the
potential (\ref{mV}) in the form
\begin{equation}\label{mVsep}
\mV(g)=g_1 R +\tilde{\mV}(g) \ .
\end{equation}
Then the contribution proportional to
$R$ given in (\ref{mVsep}) transforms
under (\ref{anWeyl}) as
\begin{eqnarray}
& &-\zeta^2 g_1\int dt d^D\bx \sqrt{g}N
B R \rightarrow
 -\zeta^2 g_1\int dt d^D\bx
N\sqrt{g}\left.[B^{\frac{2\omega-2}{\omega-D}}
R +\right.\nonumber
\\
&+& \frac{2(D-1)}{\omega-D}
B^{\frac{\omega-2+D}{\omega-D}}
\frac{\partial_i N}{N} g^{ij}\partial_j
B+ \frac{2(D-1)(2\omega-3)}
{(\omega-D)^2}
B^{\frac{2D-2}{\omega-D}}
\partial_i Bg^{ij}\partial_j B
\nonumber \\
&+&
\left.\frac{(D-1)(4-D)}{(\omega-D)^2}B^{\frac{2D-2}{\omega-D}}
\nabla_i B\nabla^i B\right] \ .
\nonumber \\
\end{eqnarray}
Using this result together with
(\ref{SFtR2kin}) and also with
$\sqrt{g}NV \rightarrow \sqrt{g}N
B^{\frac{\omega+D}{\omega-D}}V(B)$
  we find following form of the
transformed action
\begin{eqnarray}\label{genscalten}
& &\zeta^2 \int dt
d^D\bx\sqrt{g}N\left[
 K_{ij}\mG^{ijkl}K_{kl}+\frac{2}{B(\omega-D)}((1-\mu \omega)+D
 (1-\lambda))K\nabla_n B -\frac{2\mu}{B} \nabla_n
BK
+\right.\nonumber \\
&+&\frac{D}{(\omega-D)^2B^2} (1-2\mu
\omega+D(2\mu-\lambda)) \nabla_n
B\nabla_n
B+\nonumber \\
&+& B^{\frac{2D-2}{\omega-2}}
\frac{2\omega\mu(\omega-D)
-g_1(D-1)(4\omega-2-D)} {(\omega-D)^2}
\partial_i Bg^{ij}\partial_j B +
\nonumber \\
&+& \left.\frac{2\mu
(\omega-D)-2g_1(D-1)}{(\omega-D)N}
\partial_i B g^{ij}\partial_j N-g_1 B^{\frac{2\omega
-2)}{z-D}}
R(g)-\tilde{\mV}'(g,B)-B^{\frac{\omega+D}{\omega-D}}V(B)\right]
\ ,
\nonumber \\
\end{eqnarray}
where $\tilde{\mV}'$ depends explicitly
on $B$ through the relations
(\ref{OmegaB}),(\ref{RijOmega}) and
(\ref{ROmega}). It is important to
stress
 that $\omega$ is free parameter whose
 value should be determined by
 requirement that in the low energy
 limit when we can neglect the contribution
 from the
 potential $\tilde{\mV}$ and when  $\mu\rightarrow 1 \ ,
 \lambda\rightarrow 1, g_1\rightarrow
 -1$ the action (\ref{genscalten})
  flows to relativistic form
 of  scalar tensor theory
\footnote{The explicit form of the
scalar tensor  theory written in ADM
formalism is given in Appendix.}. This
requirement immediately implies that
 $\omega$ should be equal to $1$. Then the final
  form of the generalized
 scalar tensor HL gravity action
(\ref{genscalten})
 takes the form
\begin{eqnarray}\label{Sst}
S_{s.t.}&=&\zeta^2\int dt
d^D\bx\sqrt{g}N\left[
 K_{ij}K^{ij}-\lambda K^2
+\frac{2}{B(1-D)}((1-\mu )+D
 (1-\lambda))K\nabla_n B
+\right.\nonumber \\
&+&\frac{D}{(1-D)^2B^2} (1-2\mu
+D(2\mu-\lambda)) \nabla_n B\nabla_n
B+\nonumber \\
&+& \frac{2\mu(1-D)-g_1(D-1)(2-D)}
{(1-D)^2B^2}
\partial_i Bg^{ij}\partial_j B +
\nonumber \\
&+& \left.\frac{2\mu
(1-D)-2g_1(D-1)}{(1-D)N}
\partial_i B g^{ij}\partial_j N-g_1
R(g)-\tilde{\mV}'(g,B)-B^{\frac{1+D}{1-D}}V(B)\right]
\ ,
\nonumber \\
\end{eqnarray}
where the potential $\tilde{\mV}'(g,B)
$ depends on $B$ and $g_{ij}$ as
follows from the fact that it arises
from the original potential
$\tilde{\mV}$ through the anisotropic
conformal transformation. Explicitly,
the transformation (\ref{anWeyl})
implies following transformation rule
for the spatial Ricci tensor
\begin{eqnarray}
R'_{ij}&=& R_{ij}
+\frac{2(D-2)}{(1-D)^2} \frac{1}{B^2}
(\nabla_i B)(\nabla_j
B)-\frac{(D-2)}{1-D}
\frac{1}{B}\nabla_i\nabla_j B+
\nonumber \\
&+&\frac{(3-D)}{(1-D)^2}g_{ij}
\frac{\nabla_k B \nabla^k B}{B^2}
-\frac{1}{1-D}g_{ij}
\frac{\nabla_k\nabla^k B}{B} \
. \nonumber \\
\end{eqnarray}
It is also important to stress that the
covariant derivative depends on $B$ as
well. As a result the potential term
$\tilde{\mV}'$  will give the
contributions proportional to the
higher order spatial derivatives (up to
order $z$) of the field $B$ which is
the consequence of the anisotropic
scaling in $F(\tR)$ HL gravity. Note
also that is convenient to formulate
given theory in the canonical form when
we introduce the scalar field $\phi$
that is related to $B$ through the
relation
\begin{equation}\label{Bphirel}
B=e^{\Sigma \phi} \ , \quad
\Sigma=\frac{1}{\sqrt{2}}
 \frac{\sqrt{D(1-2\mu
+D(2\mu-\lambda))}}{(D-1)} \ .
\end{equation}
The action (\ref{Sst}) is the final
result of our analysis. It is useful to
compare this action (when we replace
$B$ with $\phi$ given by
(\ref{Bphirel}))
 with the
form of the scalar tensor  HL gravity
\cite{Calcagni:2009ar,Kiritsis:2009sh,
Capasso:2009fh,Suyama:2009vy,Romero:2009qs}
\begin{equation}
S=\zeta^2 \int dt d^D\bx\sqrt{g}N
[K_{ij}\mG^{ijkl}K_{kl}-
\mV(g)+\mL_{scal}] \ ,
\end{equation}
where the Lagrangian density for the
scalar field has the form
\begin{equation}
\mL=\frac{1}{2}\nabla_n \phi
\nabla_n\phi-G(g^{ij}\partial_i\phi
\partial_j\phi)-V(\phi) \ ,
\end{equation}
where $V(\phi)$ is the general
potential for the scalar field and
where $G$ is the polynomial in its
argument up to the $z-$th order. We see
that the structure of the action
(\ref{Sst}) is more complicated but it
can be again written as the polynomial
in $g^{ij}\partial_i\phi\partial_j\phi$
where now the coefficients generally
depend on $g_{ij}$ and $R_{ij}$.
 Finally,
the action (\ref{Sst}) contains the
coupling between extrinsic curvature
and the scalar field $\phi$. Note
however that the low energy limit of
(\ref{Sst}) where
$\lambda,\mu\rightarrow 1 \ ,
g_1\rightarrow -1 \ ,
\tilde{\mV}'\rightarrow 0$ takes the
form
\begin{eqnarray}\label{Sstlow}
S_{s.t.}&=&\frac{1}{(16\pi G)^2} \int
dt d^D\bx\sqrt{g}N\left[
 K_{ij}K^{ij}- K^2
-\frac{D}{(1-D)B^2} \nabla_n B\nabla_n
B+\right.\nonumber \\
&+& \left. \frac{D}{(1-D)B^{2} }
\partial_i Bg^{ij}\partial_j B +
R(g)-B^{\frac{1+D}{1-D}}V(B)\right] \ .
\nonumber \\
\end{eqnarray}
In Appendix we review the equivalence
between $F(R)$ gravity and the scalar
tensor theory written in ADM formalism
and we show that the action
(\ref{Sstlow}) exactly coincides with
the scalar tensor gravity action.

Let us conclude our result. We show
that the action for the scalar tensor
formulation of   $F(\tilde{R})$ HL
gravity is much more complicated than
in case of the $F(R)$ gravity. In other
words there is no straightforward
correspondence between $F(\tR)$
 and scalar tensor HL gravities written
 in their simplest form.
For that reason we mean that
 it is more natural to study
$F(\tR)$ HL gravities directly without
reference to their scalar tensor
images.
\\
 \noindent {\bf
Acknowledgements:}
 This work   was also
supported by the Czech Ministry of
Education under Contract No. MSM
0021622409.
 \vskip 5mm

\begin{appendix}
\section{Appendix: Equivalence between
$F(R)$ Gravity and  Scalar Tensor
Theory written in ADM
Formalism}\label{Appendix} In this
Appendix we review the well known
equivalence between $D+1$ dimensional
$F(R)$ theory of gravity and
corresponding scalar tensor theory
\footnote{For nice discussion, see
\cite{Capozziello:2005mj,Nojiri:2003ft}.}.
We perform this analysis when the
$F(R)$ action is  formulated  in ADM
formalism in order to see the relation
with the result derived  in the main
body of the paper. For that reason we
consider following form of  $F(R)$
gravity action
\begin{eqnarray}\label{SFR}
S_{F(R)}
&=&\frac{1}{(16\pi G)^2}\int dt d^D\bx
\left[ \sqrt{g}N B(
K_{ij}\mG^{ijkl}K_{kl} -R)-\nonumber
\right.
\\
&-&\left. \sqrt{g}N  V(B) -2
\sqrt{g}N\nabla_n B K  + 2
\partial_i B \sqrt{g}g^{ij}
\partial_j N \right] \ , \nonumber \\
\end{eqnarray}
where we implicitly integrated out the
scalar field $A$ so that the potential
$V(B)$ takes the same form as in
(\ref{VB}). As usual in order to find
Einstein frame form of the action
(\ref{SFR})
 we perform the conformal
rescaling of metric components
\begin{equation}
\hg_{\mu\nu}'= \Omega^2 \hg_{\mu\nu} \
 \end{equation}
that in $D+1$ decomposition takes the
form
\begin{equation}
N'=\Omega N \ , \quad N'_i=\Omega^2 N_i
\ , \quad g_{ij}=\Omega^2 g_{ij} \
\end{equation}
which is the special case of the
transformation (\ref{anWeyl}) for
$\omega=1$. Following the same analysis
as in previous section and choosing
 $\Omega=B^{\frac{1}{1-D}}$
we easily find that the kinetic term of
the $F(R)$ gravity action (\ref{SFR})
transforms as
\begin{eqnarray}\label{kinFRtr}
&&\frac{1}{(16\pi G)^2} \int dt d^D\bx
\sqrt{g}N\left[K_{ij}K^{ij}-K^2- 2
 K \nabla_n B\right]
\rightarrow \nonumber \\
&\rightarrow &\frac{1}{(16\pi G)^2}
\int dt d^D\bx \sqrt{g}N
\left[K_{ij}K^{ij}-K^2- \frac{D}{1-D}
\frac{1}{B^2}\nabla_n B\nabla_n B
\right] \
\nonumber \\
\end{eqnarray}
while the  potential term transforms as
\begin{eqnarray}\label{potFRtr}
& &\frac{1}{(16\pi G)^2}\int dt d^D\bx
\left[ -\sqrt{g}NR + 2\partial_i B
\sqrt{g} g^{ij}
\partial_j N-\sqrt{g}NV(B)\right]\rightarrow \nonumber \\
\nonumber \\
&\rightarrow & \frac{1}{(16\pi G)^2}
\int dt d^D\bx \left[-\sqrt{g}N R(g)+
\frac{D}{1-D} \sqrt{g}N \frac{1}{B^2}
\nabla_i B\nabla^i B-\sqrt{g}N
B^{\frac{1+D}{1-D}}V(B)
\right] \ .  \nonumber \\
\end{eqnarray}
Collecting (\ref{kinFRtr}) and
(\ref{potFRtr}) together  we obtain
following form of the scalar tensor
theory of gravity action
\begin{eqnarray}\label{sctegr}
& &\frac{1}{(16\pi G)^2} \int dt d^D\bx
\sqrt{g}N \left[K_{ij}K^{ij}-K^2- R(g)
-\right.\nonumber \\
&-& \left.\frac{D}{(1-D)B^2}\nabla_n
B\nabla_n B + \frac{D}{1-D}
\frac{1}{B^2} \nabla_i B\nabla^i
B-B^{\frac{1+D}{1-D}}V(B)\right] \ .
\nonumber \\
\end{eqnarray}
In order to find more familiar form of
the action (\ref{sctegr}) it is
convenient to perform the substitution
\begin{equation}\label{BphirelR}
B=\exp \frac{1}{\sqrt{2}}
\sqrt{\frac{D-1}{D}} \phi \ .
\end{equation}
Note that (\ref{Bphirel}) reduces to
(\ref{BphirelR}) in the limit when
$\mu\rightarrow 1 \ ,
\lambda\rightarrow 1$. Using
(\ref{BphirelR}) we can rewrite  the
action (\ref{sctegr}) into  the
covariant form
\cite{Capozziello:2005mj,
Nojiri:2003ft}
\begin{eqnarray}\label{actrealfin}
\frac{1}{(16\pi G)^2} \int d^{(D+1)}x
\sqrt{-\hg}
\left[R(\hg)+\frac{1}{2}\hg^{\mu\nu}
\partial_\mu
\phi\partial_\nu\phi-V'(\phi)\right] \
,
\nonumber \\
\end{eqnarray}
where
\begin{equation}
V'(\phi)=\exp
\left(\frac{1+D}{\sqrt{2}\sqrt{(D-1)D}}
\phi\right) V(\phi) \ .
\end{equation}

\end{appendix}

\newpage

 \noindent {\bf
Acknowledgements:}
 This work   was
supported by the Czech Ministry of
Education under Contract No. MSM
0021622409. \vskip 5mm



\begin{thebibliography}{20}


%
%
%
%
%
%

\bibitem{Horava:2009uw}
  P.~Horava,
\emph{``Quantum Gravity at a Lifshitz
Point,''}
  Phys.\ Rev.\  D {\bf 79} (2009) 084008
  [arXiv:0901.3775 [hep-th]].


\bibitem{Horava:2008ih}
  P.~Horava,
\emph{``Membranes at Quantum
Criticality,''}
  JHEP {\bf 0903} (2009) 020
  [arXiv:0812.4287 [hep-th]].

\bibitem{Horava:2008jf}
  P.~Horava,
\emph{``Quantum Criticality and
Yang-Mills Gauge Theory,''}
  arXiv:0811.2217 [hep-th].



\bibitem{Horava:2011gd}
  P.~Horava,
\emph{``General Covariance in Gravity
at a Lifshitz Point,''}
  arXiv:1101.1081 [hep-th].

\bibitem{Padilla:2010ge}
  A.~Padilla,
  \emph{``The good, the bad and
   the ugly .... of Horava gravity,''}
  arXiv:1009.4074 [hep-th].

\bibitem{Mukohyama:2010xz}
  S.~Mukohyama,
\emph{``Horava-Lifshitz Cosmology: A
Review,''}
  arXiv:1007.5199 [hep-th].

\bibitem{Weinfurtner:2010hz}
  S.~Weinfurtner, T.~P.~Sotiriou and M.~Visser,
\emph{``Projectable Horava-Lifshitz
gravity in a nutshell,''}
  J.\ Phys.\ Conf.\ Ser.\  {\bf 222}, 012054 (2010)
  [arXiv:1002.0308 [gr-qc]].

\bibitem{Sotiriou:2010wn}
  T.~P.~Sotiriou,
\emph{``Horava-Lifshitz gravity: a
status report,''}
  arXiv:1010.3218 [hep-th].


\bibitem{Visser:2011mf}
  M.~Visser,
\emph{``Status of Horava gravity: A
personal perspective,''}

  [arXiv:1103.5587 [hep-th]].







\bibitem{Sotiriou:2009bx}
  T.~P.~Sotiriou, M.~Visser and S.~Weinfurtner,
\emph{``Quantum gravity without Lorentz
invariance,''}
  JHEP {\bf 0910} (2009) 033
  [arXiv:0905.2798 [hep-th]].


\bibitem{Capozziello:2009nq}
S.~Capozziello, M.~De Laurentis and
V.~Faraoni, \emph{``A bird's eye view
of f(R)-gravity,''} arXiv:0909.4672
[gr-qc].

\bibitem{Nojiri:2006ri}
S.~Nojiri and S.~D.~Odintsov,
\emph{``Introduction to modified
gravity and gravitational alternative
for dark energy,''} eConf {\bf
C0602061} (2006) 06 [Int.\ J.\ Geom.\
Meth.\ Mod.\ Phys.\  {\bf 4} (2007)
115] [arXiv:hep-th/0601213].

\bibitem{Buchbinder:1992rb}
I.~L.~Buchbinder, S.~D.~Odintsov and I.~L.~Shapiro,
\emph{``Effective action in quantum
gravity,''}
{\it  Bristol, UK: IOP (1992) 413 p}

%
%
%
%


\bibitem{Chaichian:2010yi}
M.~Chaichian, S.~Nojiri,
S.~D.~Odintsov, M.~Oksanen and
A.~Tureanu, \emph{``Modified F(R)
Horava-Lifshitz gravity: a way to
accelerating FRW cosmology,''} Class.\
Quant.\ Grav.\  {\bf 27} (2010)
185021 [arXiv:1001.4102 [hep-th]]; \\

\bibitem{Carloni:2010nx}
  S.~Carloni, M.~Chaichian, S.~'i.~Nojiri, S.~D.~Odintsov, M.~Oksanen, A.~Tureanu,
 \emph{``Modified first-order
  Horava-Lifshitz gravity: Hamiltonian analysis
  of the general theory and
  accelerating FRW cosmology in power-law F(R)
  model,''}
  Phys.\ Rev.\  {\bf D82}, 065020 (2010).
  [arXiv:1003.3925 [hep-th]].


\bibitem{Nojiri:2010wj}
S.~Nojiri and S.~D.~Odintsov,
\emph{``Unified cosmic history in
modified gravity: from F(R) theory to
Lorentz
  non-invariant models,''}
arXiv:1011.0544 [gr-qc].

\bibitem{Chaichian:2010zn}
M.~Chaichian, M.~Oksanen and
A.~Tureanu, \emph{``Hamiltonian
analysis of non-projectable modified
F(R) Ho\v{r}ava-Lifshitz
  gravity,''}
Phys.\ Lett.\  B {\bf 693} (2010) 404
[arXiv:1006.3235 [hep-th]].





\bibitem{Kluson:2009rk}
  J.~Kluson,
\emph{``Horava-Lifshitz f(R)
Gravity,''}
  JHEP {\bf 0911}, 078 (2009).
  [arXiv:0907.3566 [hep-th]].





\bibitem{Kluson:2009xx}
J.~Kluson, \emph{``New Models of f(R)
Theories of Gravity,''} Phys.\ Rev.\  D
{\bf 81} (2010) 064028 [arXiv:0910.5852
[hep-th]].






\bibitem{Kluson:2010xx}
J.~Kluson, \emph{``Note About
Hamiltonian Formalism of Modified
$F(R)$ Ho\v{r}ava-Lifshitz Gravities
and Their Healthy Extension,''} Phys.\
Rev.\  D {\bf 82} (2010) 044004
[arXiv:1002.4859 [hep-th]]; JHEP {\bf
1007} (2010) 038 [arXiv:1004.3428
[hep-th]].


\bibitem{Calcagni:2009ar}
  G.~Calcagni,
\emph{``Cosmology of the Lifshitz
universe,''}
  JHEP {\bf 0909 } (2009)  112.
  [arXiv:0904.0829 [hep-th]].

\bibitem{Kiritsis:2009sh}
  E.~Kiritsis, G.~Kofinas,
\emph{``Horava-Lifshitz Cosmology,''}
  Nucl.\ Phys.\  {\bf B821 } (2009)  467-480.
  [arXiv:0904.1334 [hep-th]].

\bibitem{Capasso:2009fh}
  D.~Capasso, A.~P.~Polychronakos,
\emph{``Particle Kinematics in
Horava-Lifshitz Gravity,''}
  JHEP {\bf 1002 } (2010)  068.
  [arXiv:0909.5405 [hep-th]].

\bibitem{Suyama:2009vy}
  T.~Suyama,
\emph{``Notes on Matter in
Horava-Lifshitz Gravity,''}
  JHEP {\bf 1001 } (2010)  093.
  [arXiv:0909.4833 [hep-th]].

\bibitem{Romero:2009qs}
  J.~M.~Romero, V.~Cuesta, J.~A.~Garcia, J.~D.~Vergara,
\emph{``Conformal anisotropic mechanics
and the Horava dispersion relation,''}
  Phys.\ Rev.\  {\bf D81 } (2010)  065013.
  [arXiv:0909.3540 [hep-th]].

\bibitem{Capozziello:2005mj}
  S.~Capozziello, S.~Nojiri, S.~D.~Odintsov,
\emph{``Dark energy: The Equation of
state description versus scalar-tensor
or modified gravity,''}
  Phys.\ Lett.\  {\bf B634 } (2006)  93-100.
  [hep-th/0512118].


\bibitem{Nojiri:2003ft}
  S.~'i.~Nojiri, S.~D.~Odintsov,
\emph{``Modified gravity with negative
and positive powers of the curvature:
Unification of the inflation and of the
cosmic acceleration,''}
  Phys.\ Rev.\  {\bf D68 } (2003)  123512.
  [hep-th/0307288].

\bibitem{Arnowitt:1962hi}
  R.~L.~Arnowitt, S.~Deser, C.~W.~Misner,
 \emph{``The Dynamics of general
 relativity,''}

  [gr-qc/0405109].
\bibitem{Gourgoulhon:2007ue}
  E.~Gourgoulhon,
\emph{``3+1 formalism and bases of
numerical relativity,''}

  [gr-qc/0703035 [GR-QC]].

\bibitem{Dengiz:2011aa}
  S.~Dengiz,
\emph{``3+1 Orthogonal and Conformal
Decomposition of the Einstein Equation
and the ADM Formalism for General
Relativity,''}

  [arXiv:1103.1220 [gr-qc]].

\bibitem{Wang:2010mw}
  A.~Wang,
\emph{``$f(R)$ theory and geometric
origin of the dark sector in
Horava-Lifshitz gravity,''}
  Mod.\ Phys.\ Lett.\  A {\bf 26} (2011) 387
  [arXiv:1003.5152 [hep-th]].


\bibitem{Wang:2009azb}
  A.~Wang, D.~Wands and R.~Maartens,
\emph{``Scalar field perturbations in
Horava-Lifshitz cosmology,''}
  JCAP {\bf 1003} (2010) 013
  [arXiv:0909.5167 [hep-th]].

\bibitem{Borzou:2011mz}
  A.~Borzou, K.~Lin and A.~Wang,
\emph{``Detailed balance condition and
ultraviolet stability of scalar field
in Horava-Lifshitz gravity,''}
  JCAP {\bf 1105} (2011) 006
  [arXiv:1103.4366 [hep-th]].


\end{thebibliography}
\end{document}